\documentstyle{article}
\begin{document}

\title{Towards a Hydrodynamic Theory \\ of Infinite Neutral Nonrelativisitc Matter}
\author{Girish S. Setlur \\ Harish Chandra Research Institute \\
 Chhatnag Road, Jhusi, Allahabad, India 211 019.  } 

\maketitle

\begin{abstract}
We recast the problem of infinite neutral nonrelativistic matter interacting via U(1) gauge fields
in the hydrodynamic language. We treat the nuclei as being spinless
bosons for simplicity(for example in He4).  We write down the formal action in terms of a full set of
 independent hydrodynamic gauge invariant variables. The claim is that the results of this
 theory are nonperturbative and nuclei and electrons are treated on an equal footing.  
\end{abstract}

\section{The Theory}

Consider the action for neutral matter composed of electrons and nuclei which
we assume are all spin-zero point particles.
\begin{equation}
S = \sum_{ \sigma } \mbox{        }
\int \mbox{          } \psi^{\dagger}_{e}({\bf{x}}\sigma) 
\left( i \frac{ \partial }{ \partial t } + \frac{ \nabla^2 }{ 2m_{e} } \right)  \psi_{e}({\bf{x}}\sigma) 
+ 
 \int \mbox{          } \psi^{\dagger}_{N}({\bf{x}}) 
\left( i \frac{ \partial }{ \partial t } + \frac{ \nabla^2 }{ 2M_{ N } }
\right) \psi_{N}({\bf{x}}) 
\end{equation}
Here $ \int = \int^{ -i \beta }_{0}dt \int d^3 x $. We want this to be
invariant under gauge transformations.
\begin{equation}
\psi_{e}({\bf{x}}\sigma) \rightarrow e^{ -i e \theta({\bf{x}}) } \mbox{        }
\psi_{e}({\bf{x}}\sigma)
\end{equation}
\begin{equation}
\psi_{N}({\bf{x}}) \rightarrow e^{ i Z_{N} e \theta({\bf{x}}) } \mbox{        }
\psi_{N}({\bf{x}})
\end{equation}
Thus we have to couple to gauge fields. The invariant action with gauge fields
reads as follows.
\[
S = \int \mbox{          } \psi^{\dagger}_{e} 
\left( i \partial_{t} + e \phi - \frac{ ( - i \nabla + \frac{ e }{c}
 \mbox{   }  {\bf{A}} )^2 }{ 2m_{e} } \right)  \psi_{e} 
+  \int \mbox{          } \psi^{\dagger}_{N} 
 \left( i \partial_{t} - Z_{N} e \mbox{       }\phi - \frac{ ( -i \nabla - Z_{N}
   \frac{ e }{ c } 
 \mbox{   }  {\bf{A}} )^2 }{ 2M_{N} } \right)\psi_{N}
\]
\begin{equation}
- \frac{e}{ m_{e} c } \int {\bf{B}} \cdot \psi^{\dagger}_{e}  \mbox{
}{\bf{S}} \mbox{      } \psi_{e}
 - \int \frac{1}{4} F^2
\end{equation}
Thus in addition we have to perform the replacements,
\begin{equation}
A_{0} \rightarrow A_{0} + \partial_{t} \theta
\end{equation}
\begin{equation}
{\bf{A}} \rightarrow {\bf{A}} + \nabla \theta
\end{equation}
Now we polar decompose the fields,
\begin{equation}
\psi_{e} = e^{ i \Phi_{e} } e^{ -i \Pi_{e} } \sqrt{ \rho_{e} }  \mbox{           }
 \hspace{0.1in}; \hspace{0.1in} \psi_{N} =  e^{ -i \Pi_{N} } \sqrt{ \rho_{N} }
\end{equation}
We further postulate that $ \nabla \Phi_{e} = 0 $, this aditional phase
reflects the fermionic character of $ \psi_{e} $.
\[
S = \int \mbox{          }   \sqrt{ \rho_{e} } 
\left( C_{e} - \partial_{t} \Phi_{e}
+ \frac{ \nabla^2 - {\bf{C}}_{e}^2  + i (\nabla \cdot {\bf{C}}_{e}) + 2 i
  {\bf{C}}_{e}  \cdot  \nabla  }
{ 2m_{e} } \right) \sqrt{ \rho_{e} } 
\]
\[
+  \int \mbox{          } \sqrt{ \rho_{N} } 
 \left( - Z_{N} \mbox{       }C_{N} + \frac{ 
 \nabla^2 - Z^2_{N} {\bf{C}}_{N}^2 - i Z_{N}  ( \nabla \cdot {\bf{C}}_{N} )
 - 2 i Z_{N}  {\bf{C}}_{N}  \cdot  \nabla  }{ 2M_{N} } \right)  \sqrt{ \rho_{N} } 
\]
\[
- \frac{e}{ m_{e} c } \int {\bf{B}} \cdot   \sqrt{
  \rho_{e}({\bf{x}}\beta) }
  e^{ i \Pi_{e}({\bf{x}} \beta) } e^{ -i \Phi_{e}({\bf{x}}
  \beta) }
 \mbox{       }{\bf{S}}_{ \beta \alpha } \mbox{      } e^{ i \Phi_{e}({\bf{x}}
  \alpha) }
  e^{ -i \Pi_{e}({\bf{x}} \alpha) } \sqrt{
  \rho_{e}({\bf{x}}\alpha) }
\]
\begin{equation}
 + \int \frac{1}{2} ( {\bf{E}}^2 + {\bf{B}}^2 ) 
\end{equation}

\begin{equation}
e \phi + \partial_{t} \Pi_{e} = C_{e}
\end{equation}

\begin{equation}
\partial_{t} \Pi_{N}
 - Z_{N} e \mbox{       }\phi = - Z_{N} \mbox{       }C_{N}
\end{equation}

\begin{equation}
- \nabla \Pi_{e} +\frac{ e }{c}
 \mbox{   }  {\bf{A}} = {\bf{C}}_{e}
\end{equation}

\begin{equation}
- \nabla \Pi_{N} - Z_{N}
   \frac{ e }{ c } 
 \mbox{   }  {\bf{A}} =  - Z_{N} {\bf{C}}_{N}
\end{equation}

\begin{equation}
{\bf{E}} = - \nabla \phi - \frac{1}{ c } \frac{ \partial {\bf{A}} }{ \partial
  t }
\end{equation}

\begin{equation}
{\bf{B}} = \nabla \times {\bf{A}}
\end{equation}

\begin{equation}
- Z_{N} \nabla \Pi_{e} - \nabla \Pi_{N} =  Z_{N} {\bf{C}}_{e} - Z_{N} {\bf{C}}_{N}
\end{equation}

\begin{equation}
Z_{N} \partial_{t} \Pi_{e}  + \partial_{t} \Pi_{N} = Z_{N} C_{e} - Z_{N} C_{N}
\end{equation}

\begin{equation}
- Z_{N} \nabla \partial_{t} \Pi_{e} - \nabla \partial_{t} \Pi_{N} = 
 Z_{N} \partial_{t} {\bf{C}}_{e} - Z_{N} \partial_{t} {\bf{C}}_{N}
\end{equation}

\begin{equation}
Z_{N} \partial_{t} \nabla \Pi_{e}  + \partial_{t} \nabla \Pi_{N} = Z_{N} \nabla C_{e} - Z_{N} \nabla C_{N}
\end{equation}

\newpage

\[
S = \int \mbox{          }  
\left( \rho_{e} C_{e} - \rho_{e}\partial_{t} \Phi_{e}
- \frac{  \frac{ (\nabla \rho_{e})^2 }{ 4 \rho_{e} } + 
\rho_{e} {\bf{C}}_{e}^2  }{ 2m_{e} } \right) 
\]
\[
+  \int \mbox{          } 
 \left( - \rho_{N} \mbox{          } Z_{N} \mbox{       }C_{N} - \frac{  
 \frac{ (\nabla \rho_{N})^2 }{ 4 \rho_{N} }
 + \rho_{N}Z^2_{N} {\bf{C}}_{N}^2 }{ 2M_{N} } \right)  
\]
\[
- \frac{e}{ m_{e} c } \int {\bf{B}} \cdot \rho_{e}({\bf{x}}\alpha) 
 \mbox{       }{\bf{S}}_{ \alpha \alpha }  
- \frac{e}{ m_{e} c } \int {\bf{B}} \cdot  \sqrt{  \rho_{e}({\bf{x}} {\bar{\alpha}}) }
  e^{ i (\Pi_{e}({\bf{x}} {\bar{\alpha}})-\Pi_{e}({\bf{x}} \alpha)) }
e^{ i ( \Phi_{e}({\bf{x}} \alpha) -  \Phi_{e}({\bf{x}}{\bar{\alpha}}) ) }
 \mbox{       }{\bf{S}}_{ {\bar{\alpha}} \alpha } \mbox{      }  \sqrt{
  \rho_{e}({\bf{x}}\alpha) }
\]
\begin{equation}
 + \int \frac{1}{2} ( {\bf{E}}^2 + {\bf{B}}^2 ) 
\end{equation}

\begin{equation}
0 = \partial_{t} {\bf{C}}_{e} - \partial_{t} {\bf{C}}_{N}
 + \nabla C_{e} - \nabla C_{N}
\end{equation}

\begin{equation}
 \frac{ e }{ c } {\bf{B}} = \nabla \times {\bf{C}}_{N}
\end{equation}

\begin{equation}
e {\bf{E}} =  - \partial_{t} {\bf{C}}_{N}
 - \nabla C_{N}
\end{equation}

\begin{equation}
\partial_{t} \Pi_{e}({\bf{x}} {\bar{\alpha}})
 - \partial_{t} \Pi_{e}({\bf{x}}\alpha)
 = C_{e}({\bf{x}}{\bar{\alpha}})
 - C_{e}({\bf{x}}\alpha)
\end{equation}
We may now write in terms of the Matsubara frequencies,
\begin{equation}
\Pi_{e}({\bf{x}} {\bar{\alpha}}, t)
 - \Pi_{e}({\bf{x}}\alpha,t)
 = \sum_{ n} e^{ z_{n} t } \mbox{        }
\frac{ C_{e}({\bf{x}}{\bar{\alpha}},z_{n})
 - C_{e}({\bf{x}}\alpha,z_{n}) }{ z_{n} }
\end{equation}

\begin{equation}
{\bf{C}}_{e}({\bf{x}}\alpha,t) = {\bf{C}}_{N}({\bf{x}},t)
 - \sum_{n} \frac{ e^{  z_{n} t } }{ z_{n} } 
\left( \nabla C_{e}({\bf{x}}\alpha,z_{n}) - \nabla C_{N}({\bf{x}},z_{n}) \right)
\end{equation}

\newpage

Therefore the effective action in terms of the independent gauge invariants reads as
follows :
\[
S = \int \mbox{          }  
\left( \rho_{e} C_{e} - \rho_{e} \partial_{t} \Phi_{e}
- \frac{  \frac{ (\nabla \rho_{e})^2 }{ 4 \rho_{e} } + 
\rho_{e} ( {\bf{C}}_{N}({\bf{x}},t)
 - \sum_{n} \frac{ e^{  z_{n} t } }{ z_{n} } 
\left( \nabla C_{e}({\bf{x}}\alpha,z_{n}) - \nabla C_{N}({\bf{x}},z_{n})
\right) )^2  }{ 2m_{e} } \right) 
\]
\[
+  \int \mbox{          } 
 \left( - Z_{N} \mbox{       }\rho_{N} C_{N} - \frac{  
 \frac{ (\nabla \rho_{N})^2 }{ 4 \rho_{N} }
 + \rho_{N}Z^2_{N} {\bf{C}}_{N}^2 }{ 2M_{N} } \right)  
\]
\[
- \frac{1}{ m_{e} } \int  \nabla \times {\bf{C}}_{N} \cdot \rho_{e}({\bf{x}}\alpha) 
 \mbox{       }{\bf{S}}_{ \alpha \alpha }  
\]
\[
- \frac{1}{ m_{e} } \int  \nabla \times {\bf{C}}_{N} \cdot  \sqrt{  \rho_{e}({\bf{x}} {\bar{\alpha}}) }
  e^{ i  \sum_{ n} e^{ z_{n} t } \mbox{        }
\frac{ C_{e}({\bf{x}}{\bar{\alpha}},z_{n})
 - C_{e}({\bf{x}}\alpha,z_{n}) }{ z_{n} } }
e^{ i ( \Phi_{e}({\bf{x}} \alpha) -  \Phi_{e}({\bf{x}}{\bar{\alpha}}) ) } \mbox{      }
 \mbox{       }{\bf{S}}_{ {\bar{\alpha}} \alpha } \mbox{      }  \sqrt{
  \rho_{e}({\bf{x}}\alpha) }
\]
\begin{equation}
 + \int \frac{1}{ 2e^2 } \left[ (\partial_{t} {\bf{C}}_{N} + \nabla
 C_{N})^2  + c^2 \mbox{    }(\nabla \times {\bf{C}}_{N})^2 \right]
\end{equation}
where the fermionic phase functional is given by
\begin{equation}
\Phi_{e}([\rho];{\bf{x}}\alpha) = \sum_{ {\bf{q}}, n } \rho_{ -{\bf{q}}, -n,
  \alpha }\mbox{        }
C({\bf{q}},\alpha, n) \mbox{        }
e^{ i {\bf{q.x}} }e^{z_{n} t} 
\end{equation}

\begin{equation}
\beta z_{n} \mbox{       }C({\bf{q}},\alpha,n) = \frac{1}{ 2\left< \rho_{ {\bf{q}}
    \alpha, n } \rho_{ -{\bf{q}}
    \alpha, -n } \right>_{0} } - \frac{ (2m) \beta z^2_{n} }{ 2 N^{0}
    {\bf{q}}^2 } - \frac{ \beta {\bf{q}}^2 }{ 2 N^{0} (2m) }
\end{equation}
where  $  \left< \rho_{ {\bf{q}} \alpha, n } \rho_{ -{\bf{q}}  \alpha, -n }
\right>_{0} $ is the density-density correlation function of the
noninteracting Fermi (electron) theory.

\end{document}